\documentclass[aps,twocolumn,showpacs]{revtex4}
\usepackage[dvips]{graphicx}
\usepackage{color}
\usepackage{amsmath}
\usepackage{amsfonts}
\usepackage[dvips]{graphicx}
\usepackage{subfigure}

\begin{document}
\newcommand{\be}{\begin{equation}}
\newcommand{\ee}{\end{equation}}
\newcommand{\bea}{\begin{eqnarray}}
\newcommand{\eea}{\end{eqnarray}}
\title{Effective Polymer Dynamics of $D$-Dimensional Black Hole Interiors}

\author{Ari Peltola}
\email[Electronic address: ]{a.peltola-ra@uwinnipeg.ca} 
\author{Gabor Kunstatter}
\email[Electronic address: ]{g.kunstatter@uwinnipeg.ca} 
\affiliation{Department of Physics, The University of Winnipeg, 515 Portage Avenue, Winnipeg, Manitoba. Canada. R3B 2E9}

\begin{abstract} 
We consider two different effective polymerization schemes applied to D-dimensional, spherically symmetric black hole interiors. It is shown that polymerization of the generalized area variable alone leads to a complete, regular, single-horizon spacetime in which the classical singularity is replaced by a bounce. The bounce radius is independent of rescalings of the homogeneous internal coordinate, but does depend on the arbitrary fiducial cell size. The model is therefore necessarily incomplete. It nonetheless has many interesting features: After the bounce, the interior region asymptotes to an infinitely expanding Kantowski-Sachs spacetime. If the solution is analytically continued across the horizon, the black hole exterior exhibits asymptotically vanishing quantum-corrections due to the polymerization. In all spacetime dimensions except four, the fall-off is too slow to guarantee invariance under Poincare transformations in the exterior asymptotic region. Hence the four-dimensional solution stands out as the only example which satisfies the criteria for asymptotic flatness. In this case it is possible to calculate the quantum-corrected temperature and entropy. We also show that polymerization of both phase space variables, the area and the conformal mode of the metric, generically leads to a multiple horizon solution which is reminiscent of polymerized mini-superspace models of spherically symmetric black holes in Loop Quantum Gravity.
\end{abstract}


\maketitle

\section{Introduction}

Polymer quantization \cite{ash,halvorson} (sometimes known as Bohr quantization) provides an unitarily inequivalent alternative to Schr\"odinger quantization. It has gained importance of late because of its deep connection to Loop Quantum Gravity (LQG) \cite{ash}, and has been successfully applied to simple quantum mechanical systems \cite{ash,hl,klz} as well as mini-superspace models for gravitational systems. Investigations of loop quantum cosmology \cite{lqc1,lqc2,lqc3} and spherically symmetric black hole interiors \cite{ashtekar05,modesto06,boehmer07,pullin08,pullin08PRL,nelson} have provided strong evidence that polymer quantization may resolve the classical singularities of general relativity. 

The full polymer quantized dynamics is non-trivial even in simple models. It is therefore reassuring that there exists a limit of the polymer theory which retains some of the properties of the discretized theory while giving rise to a solvable set of dynamical equations. Recently, this effective polymerization technique was used to great effect \cite{modesto06,boehmer07,pullin08,pullin08PRL} to investigate polymer corrections to the interior of Schwarzschild black holes in 4-D spacetime. It is generally accepted that different quantization schemes (i.e. different choices of canonical variables) can lead to qualitatively different results, depending on which of the geometrical phase space variables inherit the discrete polymer structure. Corichi and Singh have postulated a reasonable set of consistency conditions which lead to a unique loop quantum gravity motivated quantization scheme in the case of homogeneous, isotropic cosmology \cite{corichi08} as well as Bianchi I cosmologies \cite{corichi09}. It is not clear whether an analoguous unique, consistent candidate exists for less symmetric situations, such as black hole interiors. One of the main purposes of this paper is to examine this question.

We investigate in detail the consequences of a non-standard choice of effective polymer dynamics for spherically symmetric black hole interiors in arbitrary spacetime dimension greater than three. The results in 4-D were presented in condensed form in \cite{pk}. Here we give the details, including results for all spacetime dimensions and an analysis of the dependence on auxiliary parameters such as the fiducial cell size introduced to regulate the integral along the non-compact spatial dimension. It proves convenient to use the formalism of generic dilaton gravity \cite{gru}, which describes the spherically symmetric sector of Einstein gravity in arbitrary spacetime dimension as well as several other 2-D models of potential interest. Since we are not tied to a particular microscopic quantum gravity theory, we will consider  polymerization schemes different from those of \cite{modesto06,boehmer07,pullin08,pullin08PRL}. We work with the parametrization of the geometric phase space variables that emerges naturally in the context of generic dilaton gravity. One of the variables is of course the dilaton itself, which, in higher dimensional Einstein gravity is proportional to the area of spheres at fixed distance from the center of symmetry. The other is the conformal mode of the 2-D metric, which is related to the conformal mode of the physical higher dimensional metric by a non-constant rescaling. 

We show first of all that polymerization of the area alone gives rise to qualitatively different spacetimes than does polymerization of both area and conformal mode. While in both quantizations the singularity is replaced by a bounce, the former leads to a single-horizon solution whereas the latter exhibits a cyclic behavior with multiple horizons, reminiscent of the results of \cite{modesto06,pullin08PRL}. As in previous works, both approaches yield a bounce radius that does depend on an extra parameter, suggesting that such models are inconsistent, as argued in \cite{corichi08}. We discuss in some detail this aspect of the model and conclude that area polymerization, in particular, while necessarily incomplete due to the dependence on fiducial scale, nonetheless yields quantum-corrected black hole spacetimes that are deserving of further study because of their intriguing generic properties. The resulting spacetimes depend on only two physical parameters, namely the black hole mass, and the bounce radius.  The latter depends on both the polymerization scale and a scale invariant version of the fiducial cell size, which needs to be fixed by heuristic arguments. It is reasonable to assume that it is roughly of the order of the Planck scale. The interior spacetime avoids the classical singularity, but the bounce is not cyclic as there is only a single bifurcative horizon. After the bounce the physical metric describes an infinitely expanding Kantowski-Sachs \cite{KS} spacetime. The polymerization thus generically drives the system into an asymptotic interior end-state that is not a small correction to the classical spacetime. The resulting scenario is reminiscent of past proposals for universe creation in black hole interiors via quantum effects \cite{frolov90,maeda}.

We also show that for $D>4$ the solutions have strange asymptotic properties in the exterior region. While the solutions are all asymptotically flat in the sense that the metrics go to the Minkowski metric at infinity, in all dimensions except four the fall-off is too slow to guarantee the finiteness of the Poincare generators. A global asymptotically flat black hole spacetime can only be constructed for $D=4$. This four-dimensional quantum-corrected black hole spacetime turns out to have fascinating properties which we review. We also present new results in 4-D for the quantum-corrected black hole temperature and entropy. As expected from  previous work (see for instance, Refs. \cite{loop_entropy,das,carlip,medved,hod}), the lowest order correction for the entropy in 4-D is logarithmic. 

The paper is organized as follows. In Section II we review classical generic dilaton gravity, including its relationship to higher dimensional spherically  symmetric Einstein gravity, Hamiltonian analysis and solution to the Hamilton-Jacobi theory. Section III describes in general the effective approach to polymer quantization. Our main results are in Section IV, which contains the polymerized (both partial and full) solutions for higher dimensional black hole interiors, as well as the extension of the solutions to the exterior. We also dedicate a subsection for a critical examination of the results obtained, including a discussion about the role of fiducial structures. Section V describes the specifics for 4-D and derives the thermodynamic properties, while Section VI ends with some conclusions and prospects for future work.

\section{Classical Theory}
Although the classical theory is well known, it is instructive to recall the main features that must be recovered in the appropriate classical limit of the polymerized theory. Furthermore, by doing so we highlight the classical scale invariance that plays an important role in the effective polymer theory. 

\subsection{Action and Solutions}
Let us begin with the action
\be \label{eq:act} S[g,\phi ] = \frac{1}{2G}\int d^2x  \sqrt{-g}\, \bigg( \phi R(g)
	+ \frac{V(\phi)}{l^2}\bigg),
\ee
which is, up to conformal reparametrizations of the metric, the most general $1+1$-dimensional, second order, diffeomorphism invariant action depending on the metric tensor $g_{\mu \nu}$ and the dilaton scalar $\phi$ \cite{gru,dil1,dil2}. In this expression, $l$ is a positive constant with a dimension of length and $G$ is the dimensionless two-dimensional Newton's constant. The quantity $S$ is dimensionless, providing we consider units in which $\hbar=1$. Specific theories are obtained by specifying the dilaton potential $V(\phi)$.

The equations implied by the action (\ref{eq:act}) are exactly solvable and obey a generalized Birkhoff theorem \cite{LMK}. The general solution can be written in ``interior Schwarzschild'' form:
\be \label{eq:ds} ds^2 = -\big[ 2lGM-j(\phi )\big]^{-1}l^2 d\phi^2+\big[ 2lGM-j(\phi )\big]dx^2,
\ee
where $j(\phi )$ satisfies
\be \label{eq:j} \frac{dj}{d\phi}=V(\phi ).
\ee
The integration constant $M$ represents the Arnowitt-Deser-Misner (ADM) mass which we take to be positive. Although more general models may be studied, in this paper we assume that $j(\phi)$ is a monotonic function such that $j(\phi)\to 0$ when $\phi\to 0$. The solution then contains precisely one Killing horizon \cite{dil2,LMK,KL} at $\phi_\text{H}$, such that:
\be
j(\phi_\text{H}) = 2lGM.
\ee

The action (\ref{eq:act}) describes a wide range of theories, each characterized by a different dilaton potential $V(\phi)$. For instance, in Jackiw-Teitelboim model \cite{JT} $V(\phi)$ is a linear function of $\phi$ whereas in Callan-Giddings-Harvey-Strominger (CGHS) theory \cite{CGHS} $V(\phi)$ is a constant. Of prime importance for the present work is that the action (\ref{eq:act}) also describes the radial sector of a spherically symmetric spacetime in $D=n+2$ dimensions. A precise correspondence can be obtained with the identifications:
{\setlength\arraycolsep{2pt}
\begin{subequations} \label{eq:correspondence}
\bea
2G&=& \frac{16\pi G^{(n+2)} n}{8(n-1){\cal \nu}^{(n)}l^n},\\
\phi&=&\frac{n}{8(n-1)}\left(\frac{r}{l}\right)^n,\\
\label{eq:correspondence3} V(\phi)&=& (n-1)\left(\frac{n}{8(n-1)}\right)^{1/n}\phi^{-1/n},
\eea
\end{subequations}%
where $G^{(n+2)}$ is the $D$-dimensional Newton's constant, $r$ is the radius of a rotationally invariant two-sphere, and
\be {\cal \nu}^{(n)}=\frac{2\pi^{(n+1)/2}}{\Gamma (\frac{1}{2}(n+1))}
\ee
is the area of the $n$-dimensional unit sphere. The dilaton has a geometrical interpretation as the area of an invariant $n$-sphere at fixed distance from the center of spherical symmetry. 

The two-dimensional metric in the action (\ref{eq:act}) is related to the physical $D$-dimensional metric by 
\be 
ds^2_\text{phys}= \frac{ds^2}{j(\phi )} + r^2d\Omega^2_n,
\label{eq:physical metric}
\ee
where $d\Omega^2_n$ is the line element of the unit $n$-sphere given in terms of the angular coordinates $\theta_i$ by:
\bea
d\Omega^2_n &=& d\theta_1^2+\sin^2(\theta_1)\,d\theta_2^2+\sin^2(\theta_1)\sin^2(\theta_2)\,d\theta_3^2+
	\cdots\nonumber\\ & &\cdots+\prod_{i=1}^{n-1}\sin^2(\theta_i)\,d\theta_n^2 .
\label{eq:unit sphere}
\eea

The physical metric has a curvature singularity at $j(\phi)=0$, and when Eqs. (\ref{eq:correspondence}) are substituted into the metric, it takes the form of the $D$-dimensional (interior) Schwarzschild solution \cite{genSchw}:
\be ds^2_\text{phys}= -\bigg(\frac{r_\text{S}^{n-1}}{r^{n-1}}-1\bigg)^{\!\!-1}dr^2 +\bigg(\frac{r_\text{S}^{n-1}}{r^{n-1}}-1\bigg)dx^2+ r^2d\Omega^2_n,
\ee
where the Schwarzschild radius $r_\text{S}$ is given by
\be 
\label{eq:Schw_r} r_\text{S}^{n-1}=\frac{16\pi G^{(n+2)}M}{n{\cal \nu}^{(n)}}.
\ee

\subsection{Hamiltonian Formulation}
We shall now proceed to the Hamiltonian formulation of the theory. We restrict to homogeneous slices in the interior. 
It is convenient to parametrize the metric in the action (\ref{eq:act}) as
\be \label{eq:adm}
ds^2 = e^{2\rho}\big( -\sigma^2dt^2+dx^2\big),
\ee
where $\rho=\rho (t)$ and $\sigma = \sigma (t)$ is the lapse function. Because the action involves an infinite integral over a spacelike coordinate $x$, we shall perform all the calculations inside a finite fiducial cell by restricting the integration over a finite interval $L_0=\int dx$. While this is a standard procedure in the existing LQG mini-superspace models, we note that it differs from our previous approach in \cite{pk} where the equations of motion were solved directly from the Lagrangian density without explicitly performing the integration. 

In terms of the parametrization (\ref{eq:adm}), and after integrating over a finite fiducial interval $L_0$, the action reads:
\be S=\int dt\Big(\Pi_\rho\dot{\rho}+\Pi_\phi\dot{\phi} +\sigma \mathcal{G}\Big),
\label{eq:action}
\ee
where a dot denotes a derivative with respect to the time coordinate $t$ and the single (Hamiltonian) constraint is:
\be 
\mathcal{G}=\frac{G}{L_0}\Pi_\phi \Pi_\rho +L_0 e^{2\rho} \frac{V}{2l^2G}\approx 0.
\label{eq:constraint}
\ee
The symbol $\approx$ denotes a weak equality in the Dirac sense: it can only be imposed after all Poisson brackets are evaluated. Since the metric is homogeneous, no timelike boundary terms are needed to complement this action. The resulting Hamiltonian equations of motion are:
{\setlength\arraycolsep{2pt}
\begin{subequations} \label{eqs:piit}
\begin{eqnarray} \Pi_\phi &=& -\frac{L_0\dot{\rho}}{\sigma G}, \\
  \label{eq:pi_rho1} \Pi_\rho &=& -\frac{L_0\dot{\phi}}{\sigma G},\\
 \dot{\Pi}_\phi &=& \frac{\sigma L_0}{2l^2G}\, e^{2\rho} \frac{dV}{d\phi },\\
  \dot{\Pi}_\rho &=& \sigma L_0 e^{2\rho} \frac{V}{l^2G}.
\end{eqnarray}
\end{subequations}}%

It is important to note that with the identifications in (\ref{eq:correspondence}), the system above is related to spherically symmetric Einstein gravity in spacetime dimension three or higher by a simple point canonical transformation. For example, for $D=4$, the Hamiltonian (\ref{eq:constraint}) can be converted to the loop quantum gravity Hamiltonian of Ref. \cite{pullin08} by the following canonical transformation:
{\setlength\arraycolsep{2pt}
\begin{subequations}
\bea
P_c&=&4\phi\,,\qquad\qquad c = -\frac{\gamma \Pi_\phi}{4} + \frac{\gamma\Pi_\rho}{16\phi},\\
P_b&=&2L_0 e^\rho \phi^{1/4}\, ,\qquad b= -\frac{\gamma}{2L_0} e^{-\rho}\Pi_\rho \phi^{-1/4},
\eea
\end{subequations}}%
where $l=l_\text{Pl}=1$ has been used. This transformation is regular for $\phi>0$.

As expected Eqs. (\ref{eq:action}-\ref{eqs:piit}) describe a parametrized Hamiltonian system with two physical phase space degrees of freedom. The Hamiltonian constraint (\ref{eq:constraint}) implies that time is embedded in one of the phase space coordinates, so that one needs to gauge fix (i.e. choose a time coordinate) in order to obtain unique evolution equations. Instead of following this procedure directly, for what follows it is useful to first obtain the classical solutions from the Hamiltonian constraint using Hamilton-Jacobi (H-J) theory \cite{landau}. We look for an H-J function $S(\phi,\rho)$ such that
{\setlength\arraycolsep{2pt}
\begin{subequations} \label{eqs:H-J_momenta}
\begin{eqnarray} \Pi_\phi &=& \frac{\partial S}{\partial \phi},\\
  \Pi_\rho &=& \frac{\partial S}{\partial \rho},
\end{eqnarray}
\end{subequations}}%
so that on the constraint surface:
\be \frac{G}{L_0}\bigg(\frac{\partial S}{\partial\phi}\bigg) \bigg(
	\frac{\partial S}{\partial\rho}\bigg) +L_0 e^{2\rho} \frac{V(\phi)}{2l^2G}=0.
\ee
This equation is trivially separable and
the ansatz $S=f(\rho )+g(\phi )$ yields the complete solution for the Hamilton-Jacobi function
\be S=-\frac{\alpha L_0}{4lG}\, e^{2\rho} +L_0\frac{j(\phi)}{\alpha lG}+C,
\ee
where $\alpha$ and $C$ are constants. The solution for $\rho$ in terms of $\phi$ is obtained from:
\be \label{eq:beta} \frac{\partial S}{\partial \alpha }=-\frac{L_0}{4lG}\, e^{2\rho} - L_0\frac{j(\phi)}{\alpha^2 lG}=-\beta ,
\ee
where $\beta$ is the constant of motion conjugate to $\alpha$. The solutions for momenta are given by
{\setlength\arraycolsep{2pt}
\begin{subequations} \label{eqs:piit2}
\begin{eqnarray} \Pi_\phi &=& \frac{\partial S}{\partial \phi } = L_0\frac{V(\phi)}{\alpha lG}, \\
  \label{eq:pi_rho2}\Pi_\rho &=& \frac{\partial S}{\partial \rho} =-\frac{\alpha L_0}{2lG}\, e^{2\rho}.
\end{eqnarray}
\end{subequations}}%

For the rest of the paper, we choose to work with positive constants $\alpha$ and $\beta$. This means that as long as $e^{2\rho}$ is positive, $\sigma$ and $\dot{\phi}$ have the same sign. This readily seen by the comparing Eqs. (\ref{eq:pi_rho1}) and (\ref{eq:pi_rho2}). In the context of black hole spacetimes, this implies that when the lapse function $\sigma$ is negative, the time evolution moves from the bifurcative black hole horizon toward the future singularity at $\phi=0$. For positive $\sigma$, the time evolution moves from the past singularity towards the bifurcative horizon. The time coordinate cannot be extended past the horizon in this formalism, since at the horizon the homogeneous initial data surface becomes null.

Given equations (\ref{eq:beta}) and (\ref{eqs:piit2}), we have now a complete solution in terms of a single arbitrary function of time and two dimensionless integration constants $\alpha$ and $\beta$. This is consistent with the fact that this is a parametrized Hamiltonian system with a two dimensional physical phase space. It is illustrative to write the solution using $\phi$ as the time coordinates since $\phi$ represents the area of the throat of the Einstein-Rosen wormhole in the extended Schwarzschild spacetime. We first solve for the lapse function,
\begin{eqnarray}
  \sigma^2 &=& \frac{4l^2\dot{\phi}^2}{\alpha^2}\, e^{-4\rho},
\end{eqnarray}
and substitute this, together with (\ref{eq:beta}), into (\ref{eq:adm}) to obtain the physical metric in  ``interior Schwarzschild'' form: 
\be ds^2 = \bigg( \frac{lG\alpha^2\beta}{L_0}-j(\phi)\bigg)^{-1}l^2 d\phi^2-
	\bigg( \frac{lG\alpha^2\beta}{L_0}-j(\phi)\bigg)\bigg( \frac{2\,dx}{\alpha}\bigg)^2.
\ee
This line element is, up to a rescaling of the spatial coordinate $x$, equivalent to (\ref{eq:ds}) with a choice $\alpha^2\beta=2L_0M$, which identifies the combination of phase space parameters that corresponds to the ADM mass. The conjugate to $M$  can be interpreted by noting that the following transformation is canonical:
\begin{subequations} \label{canonical variables}
\begin{eqnarray}
  M &=& \frac{\alpha^2\beta}{2L_0},\\
  P_M&=& \frac{2L_0}{\alpha}.
\end{eqnarray}
\end{subequations}%
It is clear from the solution, $P_M=2L_0/\alpha$ is related to a residual rescaling of the Schwarzschild ``time'' $x$. This is consistent with the Hamiltonian analysis of the exterior in which the conjugate to $M$ corresponds to the Schwarzschild time separation of the spatial slice \cite{dil1,kuchar94}. 

To summarize, Eqs. (\ref{eq:beta}) and (\ref{eqs:piit2}) provide the general solution to the theory in terms of $L_0$ and two integration constants $\alpha$ and $\beta$. $\alpha$ parameterizes the arbitrary rescaling of the coordinate $x$, which is the residual coordinate invariance after imposing homogeneity. There is also one one arbitrary function in the solution reflecting the time parametrization invariance still present in the theory. The Hamilton-Jacobi method will be used below to find the solutions in various polymerized versions of the theory.

\section{Effective Polymer Dynamics}

In the polymer representation of quantum mechanics \cite{ash,halvorson} one effectively studies
the Hamiltonian dynamics on a discrete spatial lattice. The basis states are taken to be normalizable eigenstates $|x\rangle$ 
of the position operator, such that
\bea
\langle x'|x\rangle = \delta_{x'\!\!,x}\,,
\eea
where $\delta_{x'\!\!,x}$ is the Kronecker delta and not the usual delta function. While in principle all real numbers are possible for the eigenvalues $x$, the momentum operator that generates infinitesimal translations cannot be defined on this space as a self-adjoint operator. Instead one considers the action of a finite translation operator $\hat{U}_\mu =\hat{e^{i\mu{p}}}$:
\be \hat{U}_\mu |x \rangle = |x +\mu \rangle . 
\ee
The operators $\hat{U}_\mu$ and $\hat{x}$ are self-adjoint with commutator:
\be
[\hat{x},\hat{U}_\mu]=\mu\hat{U}_\mu .
\ee
In order to construct a quantum Hamiltonian one defines a  momentum operator \cite{ash}: 
\be \label{eq:p} \hat{p} = \frac{1}{2i\mu}\big( \hat{U}_\mu -\hat{U}_\mu^\dagger \big) .
\ee
The discretization parameter $\mu >0$ is considered to be fixed so that the Hamiltonian is defined on a discrete
subset of all possible spatial points and the theory effectively lives on a lattice with 
edge length $\mu$. In principle $\mu$ can be a function of $x$, but in the following we assume that it is constant. In the limit $\mu \rightarrow 0$, Eq. (\ref{eq:p}) reduces to the standard momentum operator $\hat{p}=-i\partial_x$ and one recovers the usual Schr\"odinger quantized system \cite{corichi07}.

In many cases the full polymer theory is rather challenging to analyze but fortunately one can get interesting results by investigating the effective limit of the theory, which corresponds formally to considering the limit in which quantum effects are small, but the polymerization scale $\mu$ stays finite. In this limit the right hand side of Eq. (\ref{eq:p}) can be written in terms of a sine function of the classical momentum operator:
\be
 \label{eq:pclass} \hat{p} \to \frac{\sin(\mu p)}{\mu} .
\ee
This effective polymerization approximation is the basis for recent analyses of black hole interiors \cite{modesto06,boehmer07,pullin08,pullin08PRL}. It can be derived \cite{husain:semiclass,ding08} by studying the
action of the fully quantized operators on coherent states and expanding in the width of the states.
The end result is to simply replace the classical momentum variable 
$p$ in the classical Hamiltonian function by $\sin (\mu p)/\mu$. After the 
replacement, one studies the (semi)classical dynamics of the resulting polymer Hamiltonian 
by means of standard techniques.

\section{Polymerized Schwarzschild Interior}\label{sec:genPoly}

In order to be specific we now investigate spherically symmetric Einstein gravity in $D=n+2$ dimensions, so that the dilaton potential, given explicitly by Eq. (\ref{eq:correspondence3}), behaves as $V\propto\phi^{-1/n}$. In our considerations, 
we choose to work with a constant polymerization scale, despite the fact that in the context of loop quantum cosmology consistency with predictions requires a discreteness scale which depends explicitly on the polymerized variable(s) \cite{lqc1}. Here we choose the simplest approach that produces reasonable semiclassical behavior and leave the study of other choices for future research. Implications of non-constant polymerization scale in some LQG inspired black hole scenarios has been considered, for instance in Refs. \cite{boehmer07,nelson}.

\subsection{Partial Polymerization}

We first polymerize only the generalized area variable $\phi$. This is a somewhat ``minimalist'' approach in which we in introduce fundamental discreteness for the geometrical variable that is proportional to area in the spherically symmetric theory while leaving the coordinate dependent conformal mode of the metric continuous.  Ultimately, the real justification for this procedure is the intriguing quantum corrected black hole spacetime that emerges. As we shall see, the partial polymerization has the advantage of yielding single-horizon solutions which are not frequently encountered in semiclassical gravity.

The partially polymerized Hamiltonian constraint is
\be \label{eq:polyH}\mathcal{G} =  \frac{G}{L_0} \frac{\sin (\mu\Pi_\phi)}{\mu}\Pi_\rho
  + L_0\,e^{2\rho}\frac{V}{2l^2G}\approx 0,
\ee
and the equations of motion are given by
{\setlength\arraycolsep{2pt}
\begin{subequations} \label{eqs:eompoly}
\begin{eqnarray} \frac{\sin(\mu\Pi_\phi)}{\mu} &=& -\frac{L_0\dot{\rho}}{\sigma G}, \\
  \label{eq:dot_phi} \Pi_\rho \cos(\mu\Pi_\phi)&=& -\frac{L_0\dot{\phi}}{\sigma\ G},\\
 \dot{\Pi}_\phi &=& \frac{\sigma L_0}{2l^2G}\, e^{2\rho} \frac{dV}{d\phi },\\
  \dot{\Pi}_\rho &=& \sigma L_0\, e^{2\rho} \frac{V}{l^2G}.
\end{eqnarray}
\end{subequations}}%
The key mechanism for singularity resolution via polymerization is already evident in the above. Loosely speaking, $\dot{\phi}$ now vanishes at two turning points: the ``classical'' turning point when $\Pi_\rho=0$ and the semiclassical turning point: $\cos(\mu\Pi_\phi)=0$. The former condition will turn out to be satisfied at the horizon as expected.  In order to realize these turning points concretely and rigorously it is of course necessary to find solutions and fix a time coordinate.

As in the classical theory, we search for a solution to the corresponding Hamilton-Jacobi equation in the form 
$S=f(\rho)+g(\phi )$ to find
\be \label{eq:HJF} S=-\frac{\alpha L_0}{4lG}\, e^{2\rho}+\frac{1}{\mu}\int \arcsin 
	\bigg( \frac{L_0\mu V}{\alpha lG}\bigg)\, d\phi +C,
\ee
where $\alpha$ and $C$ are constants. As before, we take $\alpha >0$. The expressions for the momenta are now:
{\setlength\arraycolsep{2pt}
\begin{subequations}
\begin{eqnarray} \label{eq:pi1}\Pi_\phi &=& \frac{\partial S}{\partial \phi } = 
	\frac{1}{\mu}\arcsin \bigg( \frac{L_0\mu V}{\alpha lG}\bigg) , \\ \label{eq:pi2}
  \Pi_\rho &=& \frac{\partial S}{\partial \rho} =-\frac{\alpha L_0}{2lG}\, e^{2\rho}.
\end{eqnarray}
\end{subequations}%
Since the absolute value of the argument of arcsine cannot be greater than one, the polymerization imposes a condition on  $\phi$. Using the expression
(\ref{eq:correspondence3}) for $V(\phi)$:
\be \phi \geq\phi_\text{min} := c^{(n)} \bigg( \frac{L_0\mu}{\alpha lG}\bigg)^n,
\label{eq:cond} 
\ee
where 
\be c^{(n)}:=\frac{n(n-1)^{n-1}}{8}.
\ee
The minimum value of $\phi$ is located at the roots of the cosine function, as expected from (\ref{eq:dot_phi}). An inspection of the derivative $\ddot{\phi}$ verifies that this turning point is indeed a minimum.

To find the relationship between $\rho$ and $\phi$, we again differentiate $S$ with respect to $\alpha$:
\be \label{eq:beta2} \frac{\partial S}{\partial\alpha}=-\beta.
\ee
As before, $\beta$ is a constant of motion that is conjugate to $\alpha$. The explicit form of Eq. (\ref{eq:beta2}) 
depends on the given branch of the arcsine function. Eq. (\ref{eq:beta2}) can be written as
\be \label{eq:branch1} \frac{L_0}{4lG}\, e^{2\rho}+\frac{1}{\mu} I^{(n)}(\phi ) = \beta ,
\ee
where
\bea 
I^{(n)}&:=&-\epsilon\,\frac{c^{(n)}n}{a^n\alpha}\int\frac{dz}{z^n\sqrt{1-z^2}}\nonumber\\
  &\,=&-\frac{c^{(n)}n}{a^n\alpha}\int\frac{d(\mu\Pi_\phi)}{\sin^n(\mu\Pi_\phi)},
\label{eq:I}
\eea 
and we have defined
{\setlength\arraycolsep{2pt}
\begin{subequations}
\bea 
z&:=&\frac{V}{a}=\sin(\mu\Pi_\phi),\\
	\label{eq:a} a&:=&\frac{\alpha lG}{L_0\mu}.
\eea
\end{subequations}%
In Eq. (\ref{eq:I}), the value of $\epsilon=\pm 1$ depends on the given branch of $\mu\Pi_\phi$. The upper sign is valid in the branches where the cosine function is positive, which include the principal branch $(-\pi/2,\pi/2)$,  whereas the lower sign is used elsewhere. The integral $I^{(n)}$, in turn, can be evaluated in terms of $z$ in $n$-dimensions via recursive formula \cite{bryc}:
\be \int\!\!\frac{dz}{z^n\sqrt{1-z^2}}=-\frac{\sqrt{1-z^2}}{(n-1)\,z^{n-1}}
	+\frac{n-2}{n-1}\int\!\!\!\frac{dz}{z^{n-2}\sqrt{1-z^2}}.
\ee
For even values of $n$, the integral takes a rather compact form \cite{bryc}:
\be I^{(n)}\!=\epsilon\,\frac{c^{(n)}n}{a^n\alpha} \sum_{k=0}^{\frac{n}{2}-1}\frac{\Big({\textstyle{\frac{n}{2}-1} \atop \textstyle{k}}\Big)}{\textstyle{n-2k-1}}
	\bigg[ a^2\bigg(\frac{\phi}{c^{(n)}}\bigg)^{\!\!2/n}\!\!-1\bigg]^{\!(n-2k-1)/2}\!\!.
\ee
Note that in the above equation we have fixed the value of the integral by the requirement $I^{(n)}(\phi_\text{min})=0$. This in effect makes the constant $\beta$ independent of $\epsilon$ so that $e^{2\rho}$ is continuous at a branch cut of $\mu\Pi_\phi$ where $\epsilon$ changes its sign. 

\subsection{Singularity Avoidance}

In order to examine the properties of the interior solution, we now again write the physical metric using $\phi$ as the time coordinate:
\be
ds^2_\text{phys}=\frac{1}{j(\phi )}\left(\frac{-4l^2d\phi^2}{\alpha^2
	e^{2\rho}(1-V^2/a^2)}+e^{2\rho}dx^2\right)+r(\phi)^2d\Omega^2.
	\label{eq:poly metric1}
\ee
Equation (\ref{eq:poly metric1}) illustrates that as before, the solution has a horizon  when $e^{2\rho}=0$, i.e. at $\phi_\text{H}$ such that:
\be \label{eq:horizon} \frac{L_0}{4lG}\, e^{2\rho}_\text{H} = \beta - \frac{1}{\mu} I^{(n)}(\phi_\text{H} )=0.
\ee
We take $\phi_\text{H}$ to be the initial value of $\phi$ so that the corresponding initial value of $\Pi_\phi$ is
\be \mu\Pi_\phi^\text{H}:=\arcsin \left[ \frac{L_0\mu}{\alpha lG}\bigg({\frac{c^{(n)}}{\phi_\text{H}}}\bigg)^{1/n}\right].
\ee
Without loss of generality, we fix \vspace{-1.5pt} $\mu\Pi_\phi^\text{H}$ to be in the principal branch and because $\alpha$ is positive, \vspace{-0.5pt}$\mu\Pi_\phi^\text{H}$ takes its values between $(0,\pi/2)$. Note that by choosing the principal branch, and assuming again that $t$ increases toward the future, the negative values of $\sigma$ correspond to black hole hole solutions and the positive values of $\sigma$ correspond to white hole solutions, as in the unpolymerized theory.

Taking $\Pi_\phi$ as the time variable, one can deduce, generically, the following time evolution.
At $\Pi_\phi=\Pi_\phi^\text{H}$, the solution starts at the horizon. As $\Pi_\phi$ increases, $\phi$ decreases until it reaches its minimum value at $\mu\Pi_\phi =\pi/2$. At this stage, $\phi$ starts increasing again. However, when $\mu\Pi_\phi$ is in the range $(\pi/2, \pi)$, $\epsilon$ in (\ref{eq:I}) necessarily changes sign. Thus after the bounce $e^{2\rho}$ does not vanish again, and the throat area expands to $\phi\to\infty$ in finite coordinate time. However, it can be verified that the expansion takes an infinite amount of proper time. Thus, our quantization scheme has produced a solution that avoids the singularity, but does not oscillate. The time evolution of the physical conformal mode, $e^{2\rho}/j(\phi)$, is illustrated in Fig. \ref{fig:bounces} for various dimensions.
\begin{figure}[htb!]
\begin{center}
\includegraphics[scale=1.1]{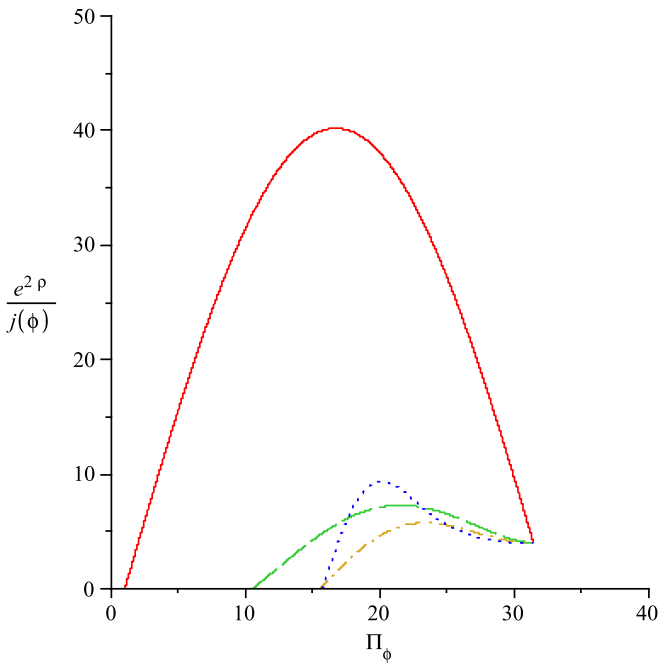}
\caption{[colour online]. Physical conformal mode plotted as a function of $\Pi_\phi$ in the models where $n=2$ (solid [red] curve), $n=3$ (dashed [green] curve),
$n=4$ (dashed and dotted [yellow] curve) and $n=10$ (dotted [blue] curve). In every model the solution begins from zero (the horizon), increases monotonically to its maximum value, and then decreases until it reaches the endpoint at $\mu\Pi_\phi=\pi$ (where $\phi\to\infty$). It can be verified that for every $n$ the solution extends toward the same endpoint, $e^{2\rho}/j(\phi)=4/\alpha^2$, which is strictly positive. Hence there is no horizon after the bounce. For numerical convenience, we have taken $l$ equal to the Planck length so that $G^{(n+2)}=l^n$, and we have used the numerical values $\alpha =l=L_0=1$, $\mu=0.1$ and $\beta=2$.} 
\label{fig:bounces}
\end{center}
\end{figure}

\subsection{Asymptotics}
In order to examine the asymptotic behavior of the solutions we now write them in terms of the areal radius $r$. A straightforward calculation reveals that
\bea
ds^2_\text{phys}&=& -\frac{dr^2}{A^{(n)}(r;M,k)\left(1-\frac{k^2}{r^2}\right)}\nonumber\\
	& &+A^{(n)}(r;M,k) \bigg(\frac{2dx}{\alpha}\bigg)^2 +r^2d\Omega^2_n,
\label{eq:xr_metric}
\eea
where we have defined
\bea \label{eq:genA}A^{(n)}&:=&\frac{n-1}{r^{n-1}}\bigg( \frac{16Gl^nM}{n^2}-I^{(n)}_\text{r}(r)\bigg)
	\nonumber\\ &\, =& \frac{n-1}{r^{n-1}}\bigg( \frac{r_\text{S}^{n-1}}{n-1}-I^{(n)}_\text{r}(r)\bigg).
\eea
In these equations:
\be
\label{eq:I1} I^{(n)}_r :=\epsilon\int \frac{r^{n-2}}{\sqrt{1-k^2/r^2}}\,dr,
\ee
and
{\setlength\arraycolsep{2pt}
\begin{subequations} \label{eq:Mk}
\bea \label{eq:M} M &:=&\frac{\alpha^2\beta}{2L_0}, \\ \label{eq:k} k &:=&\frac{L_0\mu(n-1)}{\alpha G}
	=\frac{L_0\tilde{\mu}(n-1)}{l^n\alpha G},
\eea
\end{subequations}
and $r_\text{S}$ the location of the horizon in unpolymerized theory, given by Eq. (\ref{eq:Schw_r}).
Note that in (\ref{eq:k}) $\tilde{\mu}=l^n\mu$ is the discretization scale of the physical area variable $l^n\phi\propto r^n$. In the dimensions $D=4$, $D=5$, $D=6$ and $D=7$, the explicit form of $A^{(n)}$ is, respectively,
{\setlength\arraycolsep{2pt}
\begin{subequations}
\bea A^{(2)}&=&\frac{2G^{(4)}M}{r} - \epsilon\sqrt{1-\frac{k^2}{r^2}},\\
	\label{eq:A3} A^{(3)}&=&\frac{8G^{(5)}M}{3\pi r^2} - \epsilon\sqrt{1-\frac{k^2}{r^2}}\nonumber\\
	& &\quad-\epsilon\,\frac{k^2}{r^2} \ln \bigg(\frac{r}{k}+\sqrt{\frac{r^2}{k^2}-1} \bigg),\\
	A^{(4)}&=&\frac{3G^{(6)}M}{2\pi r^3}-\epsilon\bigg( 1-\frac{k^2}{r^2}\bigg)^{3/2}\nonumber\\
	& &\quad -\epsilon\frac{3k^2}{r^2}\sqrt{1-\frac{k^2}{r^2}},\\
	A^{(5)}&=&\frac{16G^{(7)}M}{5\pi^2r^4}-\epsilon\bigg(1+\frac{3k^2}{2r^2}\bigg)\sqrt{1-\frac{k^2}{r^2}}\nonumber\\
	& &\quad -\epsilon\frac{3k^4}{2r^4}\ln \bigg(\frac{r}{k}+\sqrt{\frac{r^2}{k^2}-1} \bigg).
\eea
\end{subequations}
Hence the metric depends on two parameters, $M$ and $k$, which determine the physical properties of the solution. $M$ is the mass while $k$ gives the minimum radius of the spacetime and can be chosen independently of $M$. The fact that $k$ depends on both the fiducial length scale $L_0$ and on the scale parameter $\alpha$  raises potentially important questions about the predictive power of the model that will be addressed in detail in the next subsection. We first describe the general properties of the solution under the assumption that $k$ is microscopically small.

One can verify that the solution evolves from the horizon at $r_\text{H}$ to the minimum radius $k$ in finite proper time, and then expands to $r=\infty$ in infinite proper time (see Fig. \ref{fig:propertime} for qualitative behavior). The fact that the expansion in the interior requires infinite amount of proper time can be readily seen by integrating the proper time $\tau$ of an freely falling observer using (\ref{eq:xr_metric}), which shows that for large $r$ the proper time grows with a rate proportional to $r$.
As the radius $r$ expands in the interior, the  metric (\ref{eq:xr_metric}) approaches:
\be
ds^2_\text{phys} = -dr^2
	 +dx^2+r^2d\Omega^2_n.
\label{eq:asympt_interior}
\ee
This asymptotic interior solution does not obey the vacuum Einstein equations, but has non-vanishing stress tensor with 
{\setlength\arraycolsep{2pt}
\begin{subequations} \label{eqs:asymptT}
\bea T_x^{\, x}&=&T_r^{\, r}=-\frac{1}{8\pi G^{(n+2)}}\frac{n(n-1)}{r^2}, \\ 
	T_{\theta_i}^{\, \theta_i}&=&-\frac{1}{8\pi G^{(n+2)}}\frac{(n-1)(n-2)}{r^2}.
\eea
\end{subequations}

\begin{figure*}[htb!]
\begin{center}
\subfigure[]{\includegraphics{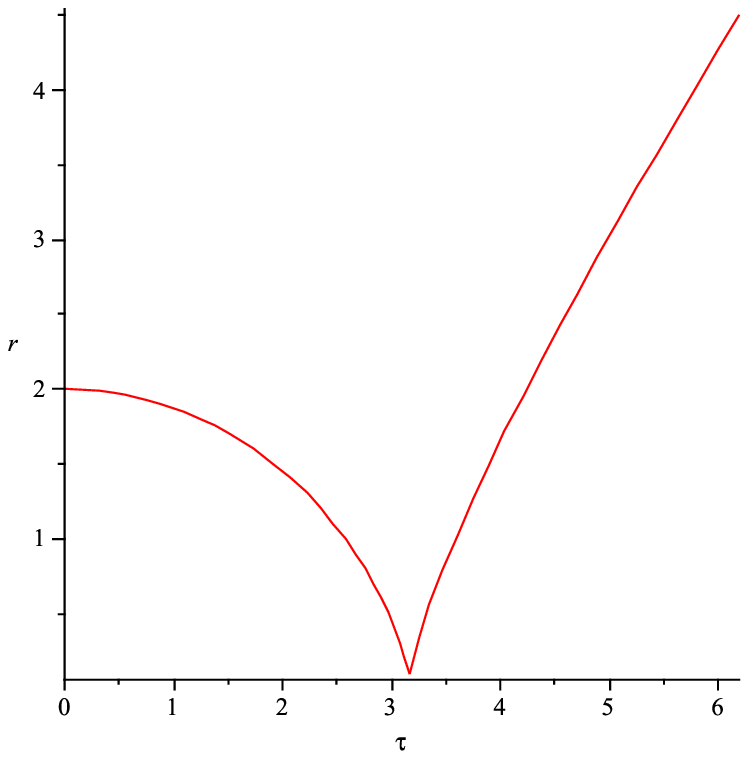}}
\hspace{2cm}
\subfigure[]{\includegraphics{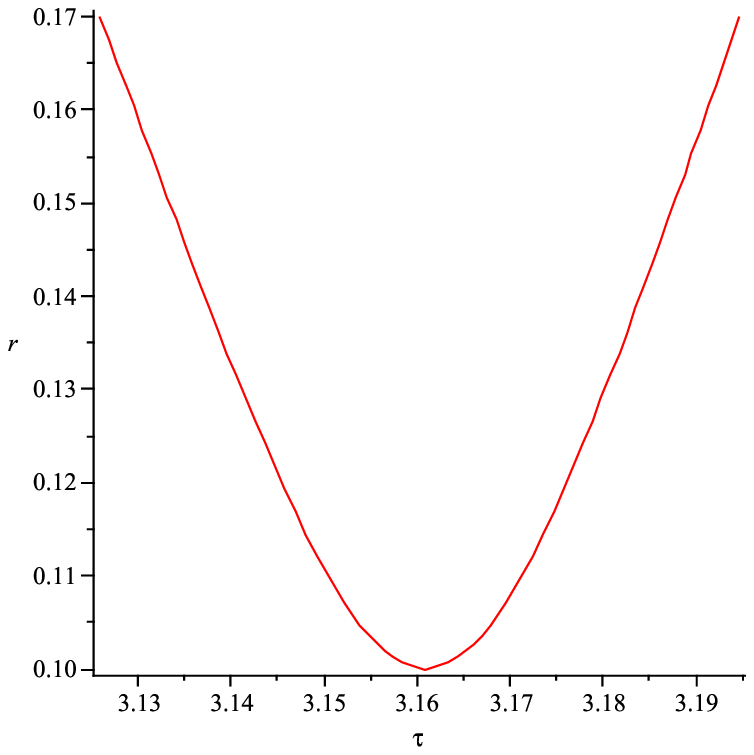}}
\caption{(a) Radial coordinate $r$ plotted as function of the proper time $\tau$ of an observer in a radial free fall (4-D). The observer falls from the horizon to the minimum radius $k$ within finite proper time. After the bounce the areal radius $r$ expands without limit, reaching infinity in infinite proper time. Despite of its appearance, there is no cusp at the bounce. (b) Close-up near the bounce radius illustrates that the curve is smooth at the bounce. In both figures $M=1$ and $k=0.1$.\label{fig:propertime}}
\end{center}
\end{figure*}

It is possible to continue the metric (\ref{eq:xr_metric}) analytically across the horizon to the exterior region. The validity of this extension is an open question given that our chosen foliation does not extend to the exterior, but the procedure seems natural in the present context as a method for constructing a complete semiclassical black hole spacetime. Similar approach has been recently used, for instance, in \cite{modesto06}. In the following, we therefore assume that the metric (\ref{eq:xr_metric}) describes the exterior region as well, with $r$ as a spacelike and $x$ as a timelike coordinate. As we shall see, in 4-D the resulting black exterior closely approximates the Schwarzschild solution whereas in higher dimensions the exterior solution differs from Einstein gravity at large (cosmological) distances.

The fact that $r=k$ is a coordinate singularity can be explicitly verified by defining a new coordinate $y$ \cite{footnote}:
\be \label{eq:coordy}
\frac{r}{k} = \cosh(y).
\ee 
The metric then takes the form:
\bea
ds^2_\text{phys} &=& -B^{(n)}(y;M,k)\,dx^2 + \frac{k^2\cosh^2(y)}{B^{(n)}(y;M,k)}\,dy^2\nonumber\\
 & & \qquad +k^2\cosh^2(y)\,d\Omega^2_n,
\label{eq:regular_metric}
\eea
where we have again absorbed $2/\alpha$ into $x$ and defined:
\bea
B^{(n)} &:=& \frac{n-1}{\cosh^{n-1}(y)}\int\cosh^{n-1}(y)\,dy\nonumber\\
	& &-\frac{16\pi G^{(n+2)}M}{n{\cal \nu}^{(n)}k^{n-1}\cosh^{n-1}(y)}.
	\label{eq:B(y)}
\eea
This coordinate system describes in a natural way one half of the complete spacetime: the exterior asymptotic region of the black hole corresponds to the limit $y\to \infty$, the horizon is located where $B^{(n)}=0$ and the minimum radius on the interior occurs at $y=0$. The asymptotic interior region corresponds to $y\to -\infty$. 

An important consequence of the form of (\ref{eq:I1}) is that the metric function $A^{(n)}$ is only dominated by the classical term $M/r^{n-1}$ in the large $r$ limit for $n=2$ ($D=4$). This is true both for the interior asymptotic region  as well as the exterior. In five spacetime dimensions, the leading term goes like
$k^2\ln(r)/r^2$, whereas in higher dimensions it is $k^2/r^2$. This is not a problem in the interior which is not asymptotically flat in any case and does not have a counterpart in the classical solution. However, in order to have a viable model for a quantum-corrected black hole, one would like the exterior to be asymptotically flat in the usual sense. In $D>4$ dimensions our semiclassical polymerized solutions do have vanishing curvature asymptotically, but they do not have a Newtonian limit as $r\to \infty$ since the Poincare generators diverge with the given fall-off conditions. The ``microscopic'' quantum corrections to the interior dynamics have a highly non-trivial effect on the global properties of the solutions. This is a somewhat surprising result. It is nonetheless reassuring that for $D=4$, which is the only case for which the corrections are well motivated by LQG considerations, the asymptotics are well behaved. We will provide more details of the four-dimensional solution in the next section.

\subsection{Auxiliary Structures}

The importance of auxiliary structures in polymerized mini-superspace models has been emphasized in Ref. \cite{corichi08} and now we turn our attention to these matters. In that article the authors propose, in the context of loop quantum cosmology, a series of well-grounded requirements which should be satisfied by a consistent cosmological model. These requirements include independence from any choice related to the auxiliary structures, e.g. the choice of coordinates or the choice of the fiducial cell, as well as the existence of a well-defined classical limit and Planckian regime. It was found that out of the existing models of flat isotropic cosmology, only \cite{lqc1} was able to satisfy all the above conditions. Recently these ideas have also been re-examined and expanded in the context of anisotropic Bianchi I cosmology \cite{corichi09}.

Analogous requirements in the context of black hole interiors, however, have not been easy to fulfill. For example, in the work of \cite{pullin08}, the quantum corrected spacetime depended on an extra integration constant which was fixed by the requirement that the bounce be symmetric. In \cite{boehmer07}, the quantization yielded a bounce that also depended on a scale invariant integration constant, so an alternative quantization scheme was proposed that yielded a bounce independent of extra parameters. In this case, however, there were quantum corrections of the (macroscopic) horizon properties, which violates another of the conditions in \cite{corichi08}. Our approach is no exception to these difficulties. The parameter $k$ given by Eq. (\ref{eq:k}) determines not only the scale at which the bounce occurs but also the curvature invariants at the bounce. It is therefore problematic that such a fundamental physical quantity depends on the fiducial length $L_0$ as well as the parameter $\alpha$. In fact it is interesting to note that the dependence of $k$ is on the ratio $L_0/\alpha$, which in the classical theory  is proportional to $P_M$ (see Eq.(\ref{canonical variables})), the momentum conjugate to the ADM mass. In the canonical theory of the full spherically symmetric spacetime $P_M$ is a Dirac observable  that is expressed as an integral over the spatial slice and invariant only under local gauge transformations (i.e. those that vanish on the boundaries of the spatial slice). It is therefore not surprising that in the present context it depends on both $L_0$ and $\alpha$, the latter, according to (\ref{eq:xr_metric}) parametrizing the residual coordinate freedom to rescale the homogeneous coordinate $x$.

In this regard, it is important to note that the particular combination $L_0/\alpha$ is invariant under rescalings of the homogeneous coordinate $x$ \cite{footnote2}. It is perhaps worth going through the argument in some detail: an examination of the metric parametrization (\ref{eq:adm}) reveals that under the rescaling $x\to b x$, the conformal mode $e^{2\rho}\to b^{-2}e^{2\rho}$, so that in the solution $\alpha\to b\alpha$ (see for example (\ref{eq:poly metric1})). Moreover, from the definition of $L_0$ it is clear that rescaling the coordinate $x$ but leaving the limits of integration unchanged results in $L_0\to b L_0$. Thus, $L_0/\alpha$ is invariant as claimed. It is also useful to examine the scaling properties of the other metric components and phase space variables. The lapse function scales as $\sigma\to b\sigma$, while the dilaton (area) $\phi$ is invariant. Applying the above information to the right hand side of the expressions for the canonical momenta, e.g. (\ref{eqs:eompoly}), one finds that they are invariant. One can also verify that the observables $M$ and $P_M$ as defined in (\ref{canonical variables}) are invariant as well. Note that for the solution analytically continued to the exterior $x$ becomes the time coordinate so $\alpha$ determines the lapse at infinity, or equivalently the relationship of the coordinate time to the proper time of an observer at infinity. This is normally taken to be unity in the Hamiltonian analysis of spherically symmetric gravity (see Ref. \cite{kuchar94}), but this need not be the case. The scale invariance of $k$ implies that once $k$ is determined for one such observer, it is in fact determined for all such observers.

This nonetheless leaves a residual (scale invariant) dependence on the initial choice of fiducial cell size, which can be interpreted in one of two ways. The first is that this dependence renders the model inconsistent, as implied in \cite{corichi08}. One would then have to find a suitable choice of variables and corresponding quantization scheme, as done in \cite{corichi08} and \cite{corichi09}, which produces a model in which physical observables do not depend on arbitrary fiducial structures.

Alternatively, one can speculate that the dependence on $L_0/\alpha$ is due to the incompleteness of the model which has not been derived directly from a microscopic theory of quantum gravity.  Certainly a more complete theory would requires a quantum description of the auxiliary structures, and these features cannot be fully captured by this simplified mini-superspace model. 
The downside of this interpretation is that one cannot make a precise prediction for the value of $k$, even if the discreteness scale for area, $\mu$, is known. However, if one is interested mostly on the qualitative aspects of semiclassical black holes, it is reasonable as well as mathematically consistent to assume that $k$ is of the order of the Planck scale, in the hope that the its actual value can be derived later from a more complete theory. While this approach is the one we adopt in the subsequent discussion, it is an important open question worthy of further study as to whether or not it can be justified at a more fundamental level.

\subsection{Fully Polymerized Theory}
For completeness we shall now present the results of the fully polymerized theory, where we introduce fundamental discreteness not only to the dilaton $\phi$ but also to the variable $\rho$. The fully polymerized Hamiltonian constraint is
\be \label{eq:fullyH}\mathcal{G} =  \frac{G}{L_0} \frac{\sin(\bar{\mu}\Pi_\rho)}{\bar{\mu}}\frac{\sin (\mu\Pi_\phi)}{\mu}
  + L_0\,e^{2\rho}\frac{V}{2l^2G}\approx 0,
\ee
where $\bar{\mu}$ denotes the dimensionless polymerization scale associated with the variable $\rho$. From this equation the time derivatives of $\rho$ and $\phi$ are obtained as
{\setlength\arraycolsep{2pt}
\begin{subequations} \label{eqs:eomfully}
\begin{eqnarray} \cos{(\bar{\mu}\Pi_\rho)}\frac{\sin(\mu\Pi_\phi)}{\mu} &=& -\frac{L_0\dot{\rho}}{\sigma G}, \\
  \frac{\sin(\bar{\mu}\Pi_\rho)}{\bar{\mu}} \cos(\mu\Pi_\phi)&=& -\frac{L_0\dot{\phi}}{\sigma G}.
\end{eqnarray}
\end{subequations}}%
The time derivatives of the momenta are unchanged.

We again search for a solution to the corresponding Hamilton-Jacobi equation in the form $S=f(\rho )+g(\phi )$ to find
\bea \label{eq:fullHJF} S&=&-\frac{1}{\bar{\mu}}\int\arcsin\bigg( \frac{L_0\bar{\mu}\alpha e^{2\rho}}{2 lG}\bigg)\,d\rho\nonumber\\ 
	& &+\frac{1}{\mu}\int\arcsin\bigg( \frac{L_0\mu V}{\alpha lG}\bigg)\, d\phi +C,
\eea
where $\alpha$ and $C$ are constants as before.
The expressions for the momenta are now:
{\setlength\arraycolsep{2pt}
\begin{subequations}
\begin{eqnarray} \label{eq:fullpi1}\Pi_\phi &=& \frac{\partial S}{\partial \phi } = 
	\frac{1}{\mu}\arcsin \bigg( \frac{L_0\mu V}{\alpha lG}\bigg) , \\
  \Pi_\rho &=& \frac{\partial S}{\partial \rho} =-\frac{1}{\bar{\mu}}\, 
   \arcsin\left(\frac{L_0\bar{\mu}\alpha e^{2\rho}}{2lG}\right).
\end{eqnarray}
\end{subequations}
Note that in the fully polymerized theory $\phi$ has the same lower bound as before, Eq. (\ref{eq:cond}), whereas $e^{2\rho}$ is bounded above:
\be
e^{2\rho} \leq \frac{2lG}{L_0\bar{\mu}\alpha}.
\ee

The solution for $\rho$ in terms of $\phi$ can be extracted from the equation
\be \label{eq:fullysol} 
I_2(\rho)+\frac{1}{\mu} I_1(\phi ) = \beta ,
\ee
where $I_1(\phi)$ is given in Eq. (\ref{eq:I}) while
\bea
	I_2(\rho)&:=& \frac{1}{2\bar{\mu}\alpha}\arcsin{\left(\frac{L_0\bar{\mu}\alpha e^{2\rho}}{2lG}\right)}\nonumber\\
  &=&-\frac{1}{2\alpha}\Pi_\rho .
\eea
Hence we have:
\be
e^{2\rho}= \frac{2lG}{L_0\alpha\bar{\mu}}\sin\big(2\bar{\mu}\alpha\beta -2\bar{\mu}\alpha I_1(\phi)/\mu\big).
\label{eq:fullrho}
\ee
All other $\phi$ dependence is unchanged, so we can write down the metric directly in terms of $r$:
\bea
ds^2_\text{phys}&=& -\frac{dr^2}{C^{(n)}(r;M,k)\left(1-\frac{k^2}{r^2}\right)}\nonumber\\
   & &+C^{(n)}(r;M,k)\left(\frac{2dx}{\alpha}\right)^2 +\,r^2d\Omega^2,
\label{eq:full_xr_metric}
\eea
where
\be \label{eq:Cn} C^{(n)}:=\frac{1}{4\bar{k} M} \frac{\,r^{n-1}_\text{S}}{\,r^{n-1}}\sin 	
	\left(4 \bar{k} M -4\bar{k} M 
	\frac{(n-1)}{r^{n-1}_\text{S}}\,I^{(n)}_r \right) ,
\ee
$M$ and $k$ are again given by (\ref{eq:Mk}) and we have defined $\bar{k}=4\bar{\mu}L_0/\alpha$ which is again scale invariant. As before the dependency on the fiducial cell size persists, so this solution is subject to the same criticism as the partially polymerized solution.

The emergence of the sine function in the fully polymerized metric gives rise to a black hole spacetime that is qualitatively different from the partially polymerized case. There will be horizons whenever the argument of the sine function in (\ref{eq:Cn}) equals multiples of $\pi$. Taking the initial value of $\mu\Pi_\rho$ to be in the principal branch, the event horizon is located at the surface where the argument of the sine is zero. (For consistency with the previous section, the initial value of $\mu\Pi_\phi$ is also taken to be in the principal branch.) After the initial condition at the event horizon has been fixed, the qualitative behavior of the solution depends on the relative magnitude of $M$, $k$ and $\bar{k}$. 

Because $I_r^{(n)}$ is an increasing function of $r$, the solution has at least one inner horizon if $4\bar{k} M >\pi$. To be more precise, the number of the inner horizons is equal to the largest positive integer $m$ which satisfies $4\bar{k} M >m\pi$.
The inner horizons may be unstable due to mass inflation \cite{mass_inflation}. Again, there is a ``bounce'' at the minimum value of the radius, $k$, as well as a new interior region, corresponding to the values $\pi/2<\mu\Pi_\phi<\pi$ (where $\epsilon=-1$). Note, however, that the new interior region consists of an infinite sequence of static and non-static regions separated by horizons. This makes it rather difficult to analyze the structure of the interior spacetime in detail, and because of this one may rather wish to construct a Reissner-Nordstr\"om type of spacetime by analytically joining together two copies of the interior. Such black hole scenarios have been recently considered in LQG \cite{modesto06,pullin08PRL}.  

We also note that there is no obvious way to extend the fully polymerized metric (\ref{eq:full_xr_metric}) to the exterior region outside the black hole. Indeed, because of the sine function in the metric, the resulting exterior spacetime would also consist of an infinite sequence of static and non-static regions separated by horizons, in a similar way as in the case of the interior region after the bounce. We leave this matter open for future consideration.

\section{4-D Schwarzschild Black Hole}
Much of the current research in singularity avoidance concentrates on four dimensional black holes, which have particular physical relevance. We therefore now review and expand on our earlier results \cite{pk} for the partially polymerized solution specifically in four spacetime dimensions. This solution has certain properties, most notably the asymptotic behaviour, which differ in crucial ways from those of the higher dimensional solutions. After describing the details of the solution, we shall derive the corrections to the black hole temperature and entropy caused by partial polymerization.

\subsection{The solution}
In terms of the radial coordinate $r$, the metric of the partially polymerized 4-D spacetime is
\bea
ds^2_\text{phys}&=& -\frac{dr^2}{\Big(\frac{2MG^{(4)}}{r} -\epsilon\sqrt{1-\frac{k^2}{r^2}}\,\Big)
	\left(1-\frac{k^2}{r^2}\right)}\nonumber\\ & &
  +\bigg(\frac{2MG^{(4)}}{r} - \epsilon\sqrt{1-\frac{k^2}{r^2}}\,\bigg)dx^2+r^2d\Omega^2,\qquad
\label{eq:four_metric}
\eea
which has a single bifurcative horizon at: 
\be
r_\text{H} := \sqrt{(2MG^{(4)})^2+k^2}.
\ee
As before, the solution evolves from the horizon at $r_\text{H}$ to the minimum radius $k$ in finite proper time, and then expands to $r=\infty$ in infinite proper time. For large $r$ in the interior, the metric approaches: 
\bea
ds^2_\text{phys} &=& -\left(1+\frac{2MG^{(4)}}{r}\right)^{\!-1}\!\!\!dr^2+\left(1+\frac{2MG^{(4)}}{r}\right)dx^2 \nonumber\\ 
	& &\qquad +\,r^2d\Omega^2.
\eea
The asymptotic interior solution has non-vanishing stress tensor with $T^{\,r}_r = T^{\,x}_x \propto -1/r^2$. Note that the angular components of the stress tensor are zero. This corresponds to an anisotropic perfect fluid that has been recently considered in a model of the Schwarzschild interior \cite{culetu}.

It is again convenient to represent the metric of the complete spacetime in terms of the coordinate $y$ of (\ref{eq:coordy}),
for which the metric takes the form:
\bea
ds^2_\text{phys} &=& -\left(\frac{\sinh(y)}{\cosh(y)}-\frac{2MG^{(4)}}{k\cosh(y)}\right) dx^2 + k^2\cosh^2(y)\times 
	\nonumber\\ & &\left[\left(\frac{\sinh(y)}{\cosh(y)}-\frac{2MG^{(4)}}{k\cosh(y)}\right)^{\!-1}\!\!\!dy^2+d\Omega^2\right] .
\eea
The exterior asymptotic region of the black hole corresponds to the limit $y\to \infty$, and the asymptotic interior region corresponds to $y\to -\infty$. The Ricci and Kretschmann scalars are nonsingular for all $y$ and vanish rapidly for large, positive $y$. A conformal diagram of the complete quantum corrected spacetime is given in Fig.(\ref{fig:conformal})
\begin{figure}[htb!]
\begin{center}
\includegraphics{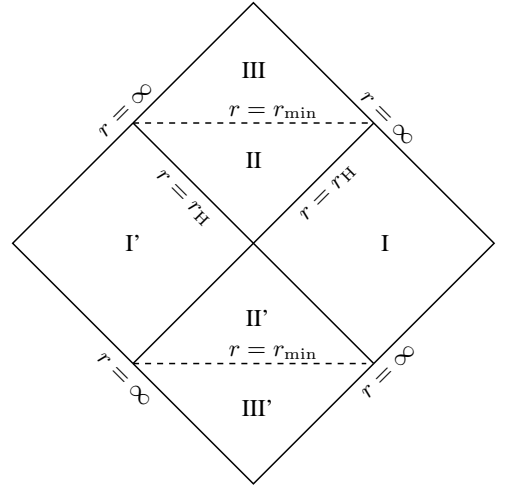}
\caption{Conformal diagram of the partially polymerized Schwarzschild spacetime. The complete spacetime includes two exterior regions (I an I'), the black hole and the white hole interior regions (II and II'), and two ``quantum corrected'' interior regions (III and III'). The classical singularity is replaced by a bounce at $r=r_\text{min}$ and subsequent expansion to $r=\infty$.}
\label{fig:conformal}
\end{center}
\end{figure}

The nonzero components of the Einstein tensor in the above coordinates are:
\bea
G_x^{\,x} &=& -\rho_1-\rho_2,\nonumber\\
G_y^{\,y} &=& -\rho_2,\nonumber\\
G_\theta^{\,\theta} &=& G_\phi^{\,\phi}= -\frac{1}{4}\rho_1,
\label{eq:einstein2}
\eea
where 
\bea
\rho_1  &=& \frac{4MG^{(4)}-2k\sinh(y)}{k^3\cosh^5(y)},\nonumber\\
\rho_2 &=&  \frac{e^{-y}}{k^2\cosh^3(y)}.
\label{eq:densities2}
\eea
For large $y$, we have $\rho_1 \to \pm 2k^2/r^4$ with the $+,-$ signs corresponding to the interior and exterior, respectively. Moreover, $\rho_2\to k^2/(2r^4)$  in the exterior, whereas it goes to $2/r^2$  in the interior. Hence the violations of the classical energy conditions are of order $k^2/r^4$ which makes them vanishingly small far from the bounce radius $r=k$. 
However, the quantum stress energy in the interior spacetime does not vanish in the limit where $k \to 0$. Instead, the asymptotic region ``pinches off'' in this limit at the curvature singularity at $r=0$, leaving behind the standard, complete but singular Schwarzschild spacetime and two disconnected, time-reversed copies of the (singular) cosmological spacetime.

\subsection{Temperature and Entropy}

One of the advantages of partial polymerization is that the polymerized metric can be naturally extended across the black hole horizon to the exterior region, giving rise to a complete, quantum-corrected black hole spacetime. The exterior region of the resulting spacetime is static and spherically symmetric, which makes it rather straightforward to obtain expressions for black hole temperature and entropy. 

The general derivation of black hole temperature in spherically symmetric 4-D spacetimes has been considered by several authors. Recent studies include Ref. \cite{hayward}, where the temperature has been obtained using the Hamilton-Jacobi tunneling method, as well as Ref. \cite{peltola}, which uses an analysis based on the Bogoliubov transformations. Defining a new radial coordinate $R$ by
\be \frac{dr}{dR}=\sqrt{1-\frac{k^2}{r^2}},
\ee
the metric in the exterior region of the partially polymerized 4-D spacetime becomes:
\be \label{eq:TR_metric}
ds^2_\text{phys}= -F(R)\,dt_\text{S}^2 +\frac{dR^2}{F(R)} +r(R)^2d\Omega^2,
\ee
where $t_\text{S}\equiv 2x/\alpha$ is a Minkowski time coordinate at asymptotical infinity and
\be F(R):=\sqrt{1-\frac{k^2}{r(R)^2}}-\frac{2G^{(4)}M}{r(R)}.
\ee
 The temperature of a corresponding macroscopic black hole can be written as \cite{hayward,peltola,visser}:
\be \label{eq:T} T=\frac{1}{4\pi}\bigg(\frac{dF}{dR}\bigg)_{r=r_\text{H}},
\ee
which represents the temperature measured by an inertial observer at asymptotic infinity. Note that this temperature can be also obtained by considering the Euclidean section of the metric (\ref{eq:TR_metric}) and requiring periodicity. 

A straightforward calculation now shows that
\bea \label{eq:T2} T&=&\frac{M}{2\pi r_\text{H}^2}\bigg( 1-\frac{k^2}{r_\text{H}^2}\bigg)^{1/2}+
	\frac{k^2}{4\pi r^3_\text{H}}\nonumber\\
	&=&\frac{1}{8\pi M}-\frac{k^2}{64\pi M^3}+\mathcal{O}\Big( \frac{k^4}{M^5}\Big),
\eea
where we have, for the sake of brevity, taken $G^{(4)}=1$. It is natural to interpret the ADM mass $M$ as the energy of the black hole so that the entropy $S$ can be obtained from the relation:
\be \frac{dS}{dM}=\frac{1}{T}.
\ee
As expected, the lowest order correction to the Bekenstein-Hawking entropy law is logarithmic: 
\be S= \frac{1}{4}A_\text{S}+\frac{\pi k^2}{2}\ln(A_\text{S}) +S_0 +\mathcal{O}(A_\text{S}^{-1}),
\ee
where
\be A_\text{S}=4\pi r_\text{S}^2= 16\pi M^2
\ee
is the area of the horizon in the unpolymerized theory, and $S_0$ is a constant. 

Note that the positive sign of the prefactor is non-standard and disagrees with the corrections arising from LQG \cite{loop_entropy}, as well with some other approaches \cite{das,carlip}. Interestingly, it does agree with the results found in \cite{hod}. The precise correspondence with \cite{hod} is obtained by choosing $k=\sqrt{2m/\pi}$, where $m$ is a natural number and it should be recalled that $k$ is given in Planck units.

\section{Conclusions}

We have presented analytic solutions to the effective polymerized dynamics of higher dimensional Schwarzschild black hole interiors using the formalism of generic dilaton gravity as a starting point. The quantum corrected solutions generically contain two independent physical parameters, the mass $M$ and the bounce radius $k$. Both are invariant under rescalings of the homogeneous coordinate, but the latter does depend on the choice of fiducial cell length $L_0$. As argued in \cite{corichi08} this suggests that further study is needed to obtain completely satisfactory singularity resolution. However, the model may turn out useful in the study of qualitative behavior of semiclassical black hole spacetimes. Under the assumption that $k$ is of the order of the Planck scale, one obtains a 4-D solution with compelling features: there is a single bifurcative horizon. On the interior the solution reaches a minimum radius $k$ before expanding into a Kantowski-Sachs type cosmological solution. The exterior black hole spacetime has quantum corrections due to the semiclassical polymerization that drop off as $O(k^2/r^2)$ and hence are very small near the horizon of macroscopic black holes. 

In higher dimensions the quantum corrections in the exterior spacetime solution do not drop off fast enough to allow a straightforward definition of the Poincare generators. They are not asymptotically flat in the usual sense. It is interesting that the polymerization seems to yield a sensible quantum corrected black hole spacetime only in 4-D. Given that the polymerization is primarily motivated by Loop Quantum Gravity, which in turn has only been formulated in four dimensions, perhaps this is to be expected. 

Finally, we note that the angular part of the spacetime metric is irrelevant when calculating the temperature of a macroscopic black hole, and because of that it should be straightforward to generalize the temperature (\ref{eq:T}) for arbitrary dimension $D$. Indeed, Euclidean arguments immediately show that (\ref{eq:T}) holds regardless of the spacetime dimension, and using (\ref{eq:genA}) we find that
\be T=\frac{n-1}{4\pi r_\text{H}},
\ee
where $r_\text{H}$ is given implicitly by the solution to (\ref{eq:horizon}). A complete treatment of the thermodynamical properties for $D>4$ must however be treated with caution, given the non-standard asymptotic behavior of the solution. This will be left for future consideration.

\section*{Acknowledgements}

We are grateful to Jon Ziprick, Jack Gegenberg, Jorma Louko, Julio Oliva and Hideki Maeda for helpful
discussions. GK also gratefully acknowledges the hospitality of the University of Nottingham, the University of New Brunswick and CECS where parts of this work were carried out. G.K. was supported in part by the
Natural Sciences and Engineering Research Council of Canada.


\begin{thebibliography}{50}

\bibitem{ash} A.~Ashtekar, S.~Fairhurst and J.~Willis, 
	``Quantum Gravity, Shadow States, and Quantum Mechanics'', 
	Class. Quant. Grav. \textbf{20}, 1031 (2003) 
	[arXiv:gr-qc/0207106].

\bibitem{halvorson} H.~Halvorson, 
	``Complementarity of Representations in Quantum Mechanics,'' 
	Studies~Hist.~Philos.~Mod.~Phys. {\bf 35}, 45 (2004) 
	[arXiv:quant-ph/0110102]. 

\bibitem{hl} V.~Husain, J.~Louko and O.~Winkler,
	``Quantum Gravity and the Coulomb Potential'',
	Phys. Rev. D \textbf{76}, 084002 (2007)
	[arXiv:0707.0273 [gr-qc]].

\bibitem{klz} G.~Kunstatter, J.~Louko and J.~Ziprick,
	``Polymer Quantization, Singularity Resolution and the $1/r^2$ Potential'',
	Phys. Rev. A \textbf{79}, 032104 (2009)
	[arXiv:0809.5098 [gr-qc]].
	
\bibitem{lqc1}
A.~Ashtekar, T.~Pawlowski and P.~Singh,
  ``Quantum Nature of the Big Bang: Improved Dynamics'',
  Phys.\ Rev.\  D {\bf 74}, 084003 (2006)
  [arXiv:gr-qc/0607039]; 

\bibitem{lqc2}
A.~Ashtekar, T.~Pawlowski, P.~Singh and K.~Vandersloot,
  ``Loop Quantum Cosmology of k = 1 FRW Models'',
  Phys.\ Rev.\  D {\bf 75}, 024035 (2007)
  [arXiv:gr-qc/0612104]; 

\bibitem{lqc3}
K.~Vandersloot,
  ``Loop Quantum Cosmology and the k = -1 RW Model'',
  Phys.\ Rev.\  D {\bf 75}, 023523 (2007)
  [arXiv:gr-qc/0612070].

\bibitem{ashtekar05}
  A.~Ashtekar and M.~Bojowald,
  ``Quantum Geometry and the Schwarzschild Singularity'',
  Class.\ Quant.\ Grav.\  {\bf 23}, 391 (2006),
  [arXiv:gr-qc/0509075].

\bibitem{modesto06}
	L.~Modesto, ``Loop Quantum Black Hole'',
	Class. Quant. Grav. \textbf{23}, 5587 (2006) [arXiv:gr-qc/0509078];
  ``Black Hole Interior from Loop Quantum Gravity'',
  Adv. High Energy Phys. \textbf{2008}, 459290 (2008) 
  [arXiv:gr-qc/0611043];
  ``Space-Time Structure of Loop Quantum Black Hole'',
  arXiv:0811.2196 [gr-qc].

\bibitem{boehmer07} C.~G.~Boehmer and K.~Vandersloot,
  ``Loop Quantum Dynamics of the Schwarzschild Interior'',
  Phys.\ Rev.\  D {\bf 76}, 104030 (2007)
  [arXiv:0709.2129 [gr-qc]]; 
  ``Stability of the Schwarzschild Interior in Loop Quantum Gravity'',
  Phys.\ Rev.\  D {\bf 78}, 067501 (2008)
  [arXiv:0807.3042 [gr-qc]]. 

\bibitem{pullin08} M.~Campiglia, R.~Gambini and J.~Pullin,
  ``Loop Quantization of Spherically Symmetric Midi-Superspaces : The Interior Problem'',
  AIP Conf.\ Proc.\  {\bf 977}, 52 (2008)
  [arXiv:0712.0817 [gr-qc]]. 

\bibitem{pullin08PRL} R.~Gambini and J.~Pullin,
  ``Black holes in Loop Quantum Gravity: the Complete Space-Time'',
  Phys.~Rev.~Lett. \textbf{101}, 161301 (2008)
  [arXiv:0805.1187].

\bibitem{nelson} W.~Nelson and M.~Sakellariadou,
	``Numerical Techniques for Solving the Quantum Constraint Equation of Generic 
	Lattice-Refined Models in Loop Quantum Cosmology'',
	Phys. Rev. D {\bf 78}, 024030 (2008)
	[arXiv:0803.4483 [gr-qc]].

\bibitem{corichi08} A.~Corichi and P.~Singh,
	``Is Loop Quantization in Cosmology Unique?'',
	Phys. Rev. D \textbf{78}, 024034 (2008)
	[arXiv:0805.0136].
	
\bibitem{corichi09} A.~Corichi and P.~Singh,
	``A Geometric Perspective on Singularity Resolution and Uniqueness in Loop Quantum Cosmology'',
	arXiv:0905.4949 [Phys. Rev. D (to be published)].

\bibitem{pk}A.~Peltola and G.~Kunstatter, 
	``A Complete, Single-Horizon Quantum Corrected Black Hole Spacetime'',
	Phys. Rev. D \emph{79}, 061501(R) (2009)
	arXiv:0811.3240 [gr-qc].
  
\bibitem{gru} D.~Grumiller, W.~Kummer and D.~V.~Vassilevich, 
	``Dilaton Gravity in Two Dimensions'', 
	Phys.~Rept. \textbf{369}, 327 (2002) 
	[arXiv:hep-th/0204253] and references therein.

\bibitem{KS} R.~Kantowski and R.~K.~Sachs, 
	``Some Spatially Homogeneous Anisotropic Relativistic Cosmological Models'',
	Journal. Math. Phys. \textbf{7}, 443 (1966).

\bibitem{frolov90} V.P. Frolov, M.A. Markov and V.F. Mukhanov, 
	``Black Holes as Possible Sources of Closed and Semiclosed Worlds'', 
	Phys. Rev. D {\bf 41}, 383 (1990); 

D.~A.~Easson, R.~H.~Brandenberger, 
	``Universe Generation from Black Hole Interiors '',
	JHEP 0106 (2001) 024 [arXiv:hep-th/0103019].

\bibitem{maeda} H.~Maeda, private communication.

\bibitem{loop_entropy}	A.~Corichi, J.~Diaz-Polo and E.~Fernandez-Borja, 
	``Loop Quantum Gravity and Planck-Size Black Hole Entropy'', 
	J. Phys. Conf. Ser. \textbf{68}, 012031 (2007) 
	[arXiv:gr-qc/0703116]. 

\bibitem{das} S.~Das, P.~Majumdar and R.~K.~Bhaduri, 
	`` General Logarithmic Corrections to Black Hole Entropy'', 
	Class. Quant. Grav. \textbf{19}, 2355 (2002) 
	[arXiv:hep-th/0111001].

\bibitem{carlip} S.~Carlip, `
	`Logarithmic Corrections to Black Hole Entropy from the Cardy Formula'', 	
	Class. Quant. Grav. \textbf{17}, 4175 (2000) 
	[arXiv:gr-qc/0005017].

\bibitem{medved} A.~J.~M.~Medved, 
	``A Comment on Black Hole Entropy or Does Nature Abhor a Logarithm?'', 
	Class. Quant. Grav. \textbf{22}, 133 (2005) 
	[arXiv:gr-qc/0406044]; 
	
	``A Follow-Up to 'Does Nature Abhor a Logarithm?' (and Apparently She Doesn't)'', 
	Class. Quant. Grav. \textbf{22}, 5195 (2005) 
	[arXiv:gr-qc/0411065].

\bibitem{hod} S.~Hod, 
	`` High-Order Corrections to the Entropy and Area of Quantum Black Holes'', 
	Class. Quant. Grav. \textbf{21}, L97 (2004) 
	[arXiv:hep-th/0405235].
	

\bibitem{dil1} D.~Louis-Martinez, J.~Gegenberg and G.~Kunstatter, 
	``Exact Dirac Quantization of All 2-D Dilaton Gravity Theories'', 
	Phys. Lett. B \textbf{321}, 193 (1994) 
	[arXiv:gr-qc/9309018].

\bibitem{dil2} J.~Gegenberg, G.~Kunstatter and D.~Louis-Martinez, 
	``Observables for Two-Dimensional Black Holes'', 
	Phys. Rev. D \textbf{51}, 1781 (1995) 
	[arXiv:gr-qc/9408015].

\bibitem{LMK} D.~Louis-Martinez and G.~Kunstatter, 
	``Birkhoff's Theorem in Two-Dimensional Dilaton Gravity'', 
	Phys. Rev. D \textbf{49}, 5227 (1994).
	
\bibitem{KL} G.~Kunstatter and J.~Louko, 
	``Transgressing the Horizons: Time Operator in Two-Dimensional Dilaton Gravity'',
	Phys.~Rev.~D \textbf{75}, 024036 (2007) 
	[arXiv:gr-qc/0608080].
	
\bibitem{JT} C.~Teitelboim,
	``Gravitation and Hamiltonian Structure in Two Spacetime Dimensions'', 
	Phys. Lett. B \textbf{126}, 41 (1983); 
	in \emph{Quantum Theory of Gravity}, ed. S.~Christensen (Adam Hilger, Bristol, 1984); 

R.~Jackiw in \emph{Quantum Theory of Gravity}, ed. S.~Christensen (Adam Hilger, Bristol, 1984);
	 
T.~Banks and L.~Susskind, 
	``Canonical Quantization of 1+1 Dimensional Gravity '',
	Int.~J.~Theor.~Phys. \textbf{23}, 475 (1984).
	
\bibitem{CGHS} C.~G.~Callan, S.~B.~Giddings, J.~A.~Harvey and A.~Strominger, 
	Phys. Rev. D \textbf{45}, R1005 (1992); 

J.~Russo, L.~Susskind and L.~ Thorlacius, 
	``End Point of Hawking Radiation'',
	Phys. Rev. D \textbf{46}, 3444 (1992);
	
S.~W.~Hawking, 
	``Evaporation of Two-Dimensional Black Holes'',
	Phys. Rev. Lett. \textbf{69}, 406 (1992).

\bibitem{genSchw} F.~R.~Tangherlini, 
	Nuovo Cimento \textbf{27}, 636 (1963); 
	
R.~C.~Myers and M.~J.~Perry, 
	``Black Holes in Higher Dimensional Space-Times'', 
	Annals Phys. \textbf{172}, 304 (1986).
	
\bibitem{landau} See, for instance, L.~D.~Landau and E.~M.~Lifshitz, 
	\emph{Mechanics} (Pergamon Press, New York, 1969).

\bibitem{kuchar94} K. Kuchar, 
	``Geometrodynamics of Schwarzschild Black Holes'', 
	Phys. Rev. D \textbf{50}, 3961 (1994) 
	[arXiv:gr-qc/9403003].


\bibitem{corichi07}
A.~Corichi, T.~Vukasinac and J.~A.~Zapata,
  ``Polymer Quantum Mechanics and Its Continuum Limit,''
  Phys.\ Rev.\  D {\bf 76}, 044016 (2007)
  [arXiv:0704.0007 [gr-qc]].

\bibitem{husain:semiclass} V.~Husain, O.~Winkler, 
	``Semiclassical States for Quantum Cosmology'', 
	Phys. Rev. D \textbf{75} (2007) 024014 
	[arXiv:gr-qc/0607097].
	
\bibitem{ding08} Ding Wang, R. B. Zhang, Xiao Zhang, 
	``Quantum Deformations of Schwarzschild and Schwarzschild-de Sitter Spacetimes'', 
	Class. Quant. Grav. \textbf{26}, 085014 (2009) 
	[arXiv:0809.0614 [hep-th]].

\bibitem{bryc} Yu.~A.~Brychkov, O.~I.~Marichev and A.~P.~Prudnikov, \emph{Tables of Indefinite
	Integrals}, translated by G.~G.~Gould (Gordon and Breach Science Publishers, New York, 1989).

\bibitem{footnote} G. K. is grateful to Julio Oliva and Hideki Maeda for suggesting these coordinates.

\bibitem{footnote2} This was also pointed out in \cite{boehmer07} concerning their analoguous parameter $p_b^{(0)}$.
	
\bibitem{mass_inflation} E.~Poisson and W.~Israel, ``Internal Structure of Black Hole'', Phys. Rev. D \textbf{41} 1796 (1990); 	``Inner-Horizon Instability and Mass Inflation in Black Holes'', Phys. Rev. Lett. \textbf{63}, 1663 (1989); ``Eschatology of the Black Hole Interior'', Phys. Lett. B \textbf{233}, 74 (1989). 

\bibitem{culetu}H. Culetu, 
	``On The Black Hole Interior Spacetime'', 
	Int. J. Mod. Phys. A \textbf{24}, 1593 (2009) 
	[arXiv:hep-th/0701255].


\bibitem{hayward} S.~A.~Hayward, R.~Di.~Criscienzo, L.~Vanzo, M.~Nadalini and S.~Zerbini, 
	``Local Hawking Temperature for Dynamical Black Holes'', 
	Class. Quantum Grav. \textbf{26}, 062001 (2009)
	[arXiv:0806.0014 [gr-qc]].

\bibitem{peltola} A.~Peltola, 
	``Local Approach to Hawking Radiation'', 
	Class. Quant. Grav. \textbf{26}, 035014, (2009) 
	[arXiv:0807.3309 [gr-qc]].

\bibitem{visser} M.~Visser, 
	``Essential and Inessential Features of Hawking Radiation'', 
	Int. J. Mod. Phys. D \textbf{12}, 649 (2003) 
	[arXiv:hep-th/0106111].

\end{thebibliography}
\end{document}